\documentclass[preprint2]{aastex}

\textheight=\baselinestretch\textheight
\ifdim\textheight>25.2cm\textheight=25.0cm\fi
\topskip\baselineskip
\maxdepth\baselineskip
\usepackage[bookmarks=false]{hyperref}
\usepackage{epsfig}
\usepackage{amsmath}
\usepackage{longtable}
\usepackage{lscape}

\newbox\grsign \setbox\grsign=\hbox{$>$}
\newdimen\grdimen \grdimen=\ht\grsign
\newbox\laxbox \newbox\gaxbox
\setbox\gaxbox=\hbox{\raise.5ex\hbox{$>$}\llap
     {\lower.5ex\hbox{$\sim$}}}\ht1=\grdimen\dp1=0pt
\setbox\laxbox=\hbox{\raise.5ex\hbox{$<$}\llap
     {\lower.5ex\hbox{$\sim$}}}\ht2=\grdimen\dp2=0pt

\def\lax{\mathrel{\copy\laxbox}}
\def\simless{\lax}

\shorttitle{Detection of Low-Amplitude $\delta$ Sct Stars in TAOS 2-Year Data}
\shortauthors{D.-W. Kim et al.}

\begin{document}

\title{The TAOS Project Stellar Variability I. \\
Detection of Low-Amplitude $\delta$ Scuti Stars}

\author
{D.-W. Kim\altaffilmark{1,2,3},
P. Protopapas\altaffilmark{1,2}, C. Alcock\altaffilmark{1}, Y.-I. Byun\altaffilmark{3},  J. Kyeong\altaffilmark{4}, 
B.-C. Lee\altaffilmark{4}, N. J. Wright\altaffilmark{1},
T. Axelrod\altaffilmark{5}, F. B. Bianco\altaffilmark{1,6},
W.-P. Chen\altaffilmark{7},  N. K. Coehlo\altaffilmark{8}, K. H. Cook\altaffilmark{9}, R. Dave\altaffilmark{2}, S.-K. King\altaffilmark{10},
T. Lee\altaffilmark{10},  M. J. Lehner\altaffilmark{10,6,1},  H.-C. Lin\altaffilmark{7}, 
S. L. Marshall\altaffilmark{9,11},
R. Porrata\altaffilmark{12}, J. A. Rice\altaffilmark{8}, M. E. Schwamb\altaffilmark{13},
J.-H. Wang\altaffilmark{7,10}, S.-Y. Wang\altaffilmark{10}, C.-Y. Wen\altaffilmark{10} and Z.-W. Zhang\altaffilmark{7,10}
}

\altaffiltext{1}{Harvard Smithsonian Center for Astrophysics, Cambridge, MA 02138}
\altaffiltext{2}{Initiative in Innovative Computing,  School of Engineering and Applied Sciences, Harvard, Cambridge, MA 02138}
\altaffiltext{3}{Department of Astronomy, Yonsei University, Seoul, South Korea 120-749}
\altaffiltext{4}{Korea Astronomy \& Space Science Institute, Daejeon 305-348, Korea}
\altaffiltext{5}{Steward Observatory, 933 North Cherry Avenue, Room N204  Tucson AZ 85721}
\altaffiltext{6}{Department of Physics and Astronomy, University of  Pennsylvania, 209 South 33rd Street, Philadelphia, PA 19104}
\altaffiltext{7}{Institute of Astronomy, National Central University, No. 300, Jhongda Rd, Jhongli City, Taoyuan County 320, Taiwan}
\altaffiltext{8}{Department of Statistics, University of California Berkeley,  367 Evans Hall, Berkeley, CA 94720}
\altaffiltext{9}{Institute for Geophysics and Planetary Physics, Lawrence  Livermore National Laboratory, Livermore, CA 94550}
\altaffiltext{10}{Institute of Astronomy and Astrophysics, Academia Sinica. P.O. Box 23-141, Taipei 106, Taiwan}
\altaffiltext{11}{Kavli Institute for Particle Astrophysics and Cosmology,  2575 Sand Hill Road, MS 29, Menlo Park, CA 94025}
\altaffiltext{12}{Department of Astronomy, University of California Berkeley,  601 Campbell Hall, Berkeley CA 94720}
\altaffiltext{13}{Division of Geological and Planetary Sciences, California  Institute of Technology, 1201 E. California Blvd., Pasadena, CA 91125 }




\begin{abstract}

   We analyzed data accumulated during 2005 and 2006 by the Taiwan-American Occultation Survey (TAOS)  in order to detect short-period variable stars (periods of $\simless$ 1 hour) such as $\delta$ Scuti. 
   TAOS is designed for the detection of stellar occultation by
small-size Kuiper Belt Objects (KBOs) and is operating four 50cm telescopes at an effective cadence of 5Hz.
The four telescopes simultaneously  monitor the same patch of the sky in order  to reduce false positives.
To detect short-period variables, we used the Fast Fourier Transform algorithm (FFT)
inasmuch as the data points in TAOS light-curves are evenly spaced.
Using FFT, we found 41 short-period variables with amplitudes smaller than a few hundredths of a magnitude and periods of  about an hour,
which suggest that they are low-amplitude $\delta$ Scuti stars (LADS).
The light-curves of TAOS $\delta$ Scuti stars are accessible online at the
Time Series Center website (http://timemachine.iic.harvard.edu).

\end{abstract}


\section{INTRODUCTION}

$\delta$ Scuti stars (hereinafter, $\delta$ Sct stars) are pulsating variables inside the classical instability strip  and on or close to the main-sequence.
They are typically placed at the lower on the instability strip than  RR Lyrae stars or Cepheids and
thus they are fainter than RR Lyrae stars or Cepheids.
Their spectral types are between A and late F.
Their periods are between $\sim$0.02 days and $\sim$0.25 days, which is relatively shorter
than other types of variables (e.g. $\gamma$ Dor; \citet{Henry2001}). 
Based on these characteristics, $\delta$ Sct stars can be separated  from other types of variable stars
such as RR Lyrae, $\beta$ Cepheid, $\gamma$ Dor etc \citep{Breger2000}.

The majority of the $\delta$ Sct stars are low-amplitude $\delta$ Sct stars (LADS) with amplitudes from a milli-magnitude  to a few tens of milli-magnitude.
LADS are mainly non-radial p-mode pulsators \citep{Breger2000}.
Another subgroup of $\delta$ Sct stars is the high-amplitude $\delta$ Sct stars (HADS),
whose amplitudes are bigger than  $\sim$0.3 magnitude. HADS are radial pulsators \citep{Breger2000, Rodriguez1996}.
In addition to LADS and HADS, there is
another interesting type  pulsation star called SX Phe variable stars, which exhibit a 
type of pulsation similar to the $\delta$ Sct stars.
They are relatively old and evolved Population II stars, whereas most of the $\delta$ Sct stars are  Population I stars.
Stellar evolutionary theory is not yet successful at explaining these SX Phe variable stars \citep{Rodriguez2000.2}.
Most of the SX Phe show similar properties with HADS such as high amplitude and short period.
More detailed review of $\delta$ Sct stars is presented in \citet{Breger2000} \citep{Breger2000} and references therein.

Because of their great number of radial and non-radial modes, 
it is known that $\delta$ Sct stars are suitable for asteroseismology research, which enables study of stellar
interior structures \citep{Brown1994}. For a better understanding of
pulsating $\delta$ Sct stars and thus stellar structure,
several authors studied $\delta$ Sct stars and detected their
multiple frequencies of pulsation using either ground-based observations or space-based observations \citep{Breger2002, Ripepi2003, Breger2005, Buzasi2005, Bruntt2007a, Pribulla2008}.
Due to the better photometric precision, space-based observations data show better
results on the analysis of multiple frequencies than ground-based observation  data \citep{Bruntt2007a, Pribulla2008}.
However, some authors have pointed out that ground-based observations using multiple-site telescopes are
still valuable because, with a baseline longer than space-based observations, 
they are useful for detecting long-period pulsation (for more details, see \citealt{Breger2005, Bruntt2007a} and references therein).
Moreover, by parameterizing the amplitude ratio and the phase differences in different filters (e.g. $ubvy$),
it is possible to derive the spherical harmonic degree, $l$ \citep{Garrido1990, Balona1999, Moya2004},
which is an important parameter for the asteroseismology studies.
Therefore ground-based telescopes that are more feasible for multiple-site and
multiple-filter observations (e.g. Delta Scuti Network, \citealt{Zima2002}) are nonetheless useful
for the identification of pulsation modes and thus for the study of interior structures.

Another interesting feature of $\delta$ Sct stars is that some of them
show period and amplitude variations \citep{Breger1998, Breger2000.313, Arentoft2001}.
The period variation (1/P) dP/dt, based on observations, is about 10$^{-7}$ per year for
both period increases and decreases with equal distribution. On the other hand, theoretical models give 
ten times smaller period variation than observed; they also predict that period increases should be dominant over period decreases \citep{Breger1998}. Amplitude variations and time scales of the variations
are different from star to star, ranging from a few milli-magnitudes to several tens of milli-magintudes and from a few tens of days to a few hundreds of days \citep{Arentoft2001}.
These period and amplitude variations are thought to be caused not by evolutionary effects
but by some other mechanism (e.g. light-time effect because of the orbital motion in binaries or
nonlinear mode interactions). However, the true origin of the
variations is still unknown.
For more details, see \citet{Breger1998} and references therein. 

\citet{McNamara2007} investigated HADS in the Large Magellanic Cloud (LMC) and
their period-luminosity (P-L) relation to test if they can
be used as the standard distance candles.
They found that the distance modulus for LMC derived using $\delta$ Sct stars is consistent with
the distance moduli for LMC derived using RR Lyrae and Cepheids,
which implies the P-L relation of $\delta$ Sct stars can help to determine
distances of {\em long-distance} objects such as objects in the LMC.

In this paper, we present the detection of 41 $\delta$ Sct candidate stars
from the Taiwan-American Occultation Survey (TAOS) data accumulated during 2005 and 2006 observation (hereinafter, TAOS $\delta$ Sct stars).
Among the 41 detections, there is one previously known `suspected variable' star, NSV 3816, from the Suspected Variable stars and Supplement \citep{Samus2009} (no period or type is provided in that catalog). 
The rest of the 40 TAOS $\delta$ Sct stars are newly detected by this study.
Only 14 of the detected TAOS $\delta$ Sct stars have spectral types. 
Twelve of those have spectral types  from A to F, which are typical for $\delta$ Sct stars. 
The remaining  two have B8 and G5 spectral types, which are peculiar spectral types.
Using spectroscopic instruments -BOES  \citep{Kim2007} and FAST  \citep{Fabricant1998}-, we obtained spectra  for  those two stars. 
As a result we found that the B8 star is an A5 star and the G5 star is an F0 star.
Even though the rest of the detected stars do not have spectral type information, 
their  low amplitudes, short periods and morphologies of light-curves strongly suggest that they are LADS.

In Section \ref{sec:TAOS_Delta_Scuti}, we present a TAOS overview, data reduction processes and the detection algorithm 
we used to detect $\delta$ Sct stars in TAOS 2-year data.
We provide a list of the detected TAOS $\delta$ Sct stars and their physical parameters
(e.g. magnitude, period, amplitude, spectral type, etc) in Section \ref{sec:DetectionResults}. 
In Section \ref{sec:Conclusion}, we present summaries.

\section{TAOS  $\delta$ Sct Stars}
\label{sec:TAOS_Delta_Scuti}

\subsection{TAOS overview}

TAOS aims to detect stellar occultations caused by small-sized Kuiper Belt Objects (KBOs)
at a distance of Neptune's orbit or beyond \citep{Alcock2003, Chen2007, Lehner2009}.
Because of the short duration ($<1$ sec)
and the rareness of occultation events, TAOS monitors several hundreds of stars in a wide field of view (3 deg${}^2$)
with a high sampling rate.
To reduce false positives,
TAOS uses four $50\,$cm telescopes which simultaneously monitor the same patch of the sky.
Due to the high sampling rate, TAOS data is also useful for
detecting short-period variable stars such as $\delta$ Sct stars.
Moreover, TAOS telescopes keep monitoring the same field up to 1.5 hours and can thus obtain
full-phase light-curves of variable stars whose periods are shorter than 1.5 hours.

To detect such short-period variable stars, we analyzed TAOS  data accumulated during 2005 and 2006.
The dataset consists of 117 TAOS observation fields, which cover 351 $\text{deg}^2$ of the sky.
It consists of $\sim$200 {\em runs}, where a run is a set of multiple
(two or three\footnote{During 2005 and 2006 observational season, one of the four TAOS telescopes was not operational.})
telescope observations for a given field and a given date.
Note that the TAOS telescopes occasionally visit the same observation field multiple times
according to the telescopes' observation schedules,
which enables detecting the same variable stars multiple times.
In such case, we are able to derive multiple frequencies of the stars
as explained in Section \ref{sec:ListDeltaScuti}.

\subsection{Data Reduction}

To detect periodic signals, we analyzed the light-curves generated by
the TAOS photometry pipeline \citep{Zhang2009}.
The pipeline was developed by the collaboration to extract light-curves of each star from {\em zipper images}.
The {\em zipper images} are generated by the unique telescope operation mode
called {\em zipper mode} which was developed  in order to achieve high-speed photometry \citep{Lehner2009}.

After obtaining the light-curves using the TAOS photometry pipeline, we applied further cuts to the light-curves.
Some of the individual measurements  are flagged as invalid. This happens when the star moves out of
the field of view because of temporary telescope vibrations or tracking error, thus yielding no photometrical measurements.
We therefore applied a B-spline \citep{Boor1978} and replaced the flagged measurements with values
interpolated from the spline fit. After the interpolation 
process,  in order to increase the signal-to-noise ratio (SNR),  
we binned each light-curve using a 50 point window (10 sec).
During the binning process, we used the average time of the 50 data points as the time of the binned data.

We then removed the systematic variations that are common across light-curves of the same run.
Such systematic variations, which we call {\em trends}, could be caused by airmass,
temporary telescope vibrations, noise in  CCD images, etc.
To remove such trends, we applied the Photometric DeTrending algorithm (PDT, \citealt{Kim2009}) to each individual run.
PDT first calculates the correlation between whole light-curves 
as a measure of similarity between light-curves.
PDT then uses the hierarchical clustering algorithm \citep{Jain1999} to group similar light-curves together
and  determines one {\em master-trend} per group by summing weighted light-curves in the group.
Using the determined master-trends, PDT finally removes trends from each individual
light-curve by minimizing the residual between the master-trends and the light-curve.
For more details about PDT, see \citet{Kim2009}.

\begin{figure}[tbp]
\begin{center}
       \includegraphics[width=0.45\textwidth]{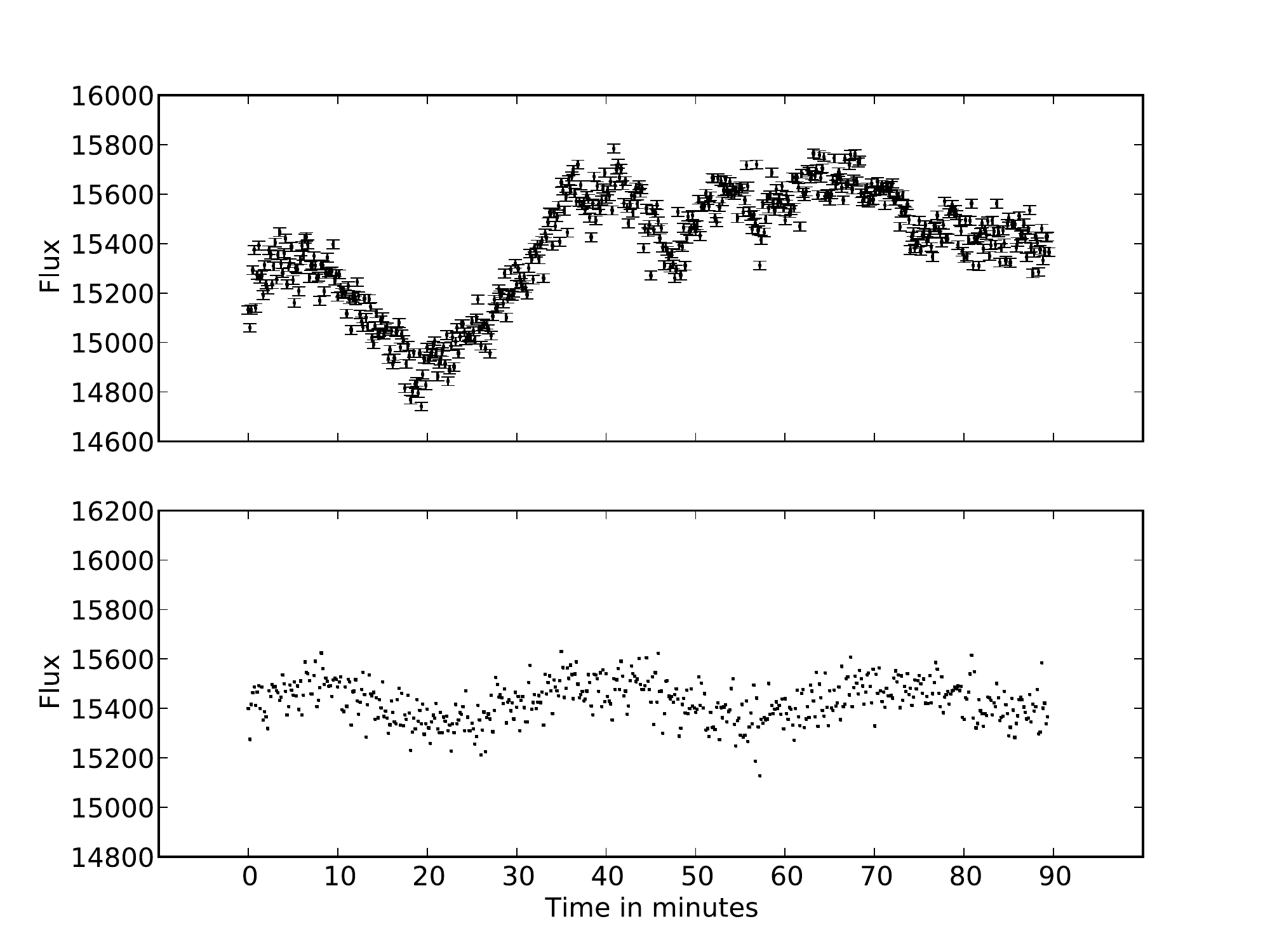}
\end{center}
    \caption{An example light-curve of a TAOS $\delta$ Sct star.
    X-axis is time in minutes and y-axis is flux.
    The top panel is the light-curve before detrending and the bottom panel is the light-curve after detrending.
    The periodic signals contaminated by unstable weather (e.g. moving clouds) are successfully
    recovered after detrending. We show the errors for each photometric measurement of the raw light-curve.}
    \label{fig:LC_Example}
\end{figure}

Figure \ref{fig:LC_Example} shows an example of a TAOS $\delta$ Sct star's light-curve 
before and after detrending process. X-axis is time in minutes and y-axis is flux.
As the figure shows, periodic signals are clearly recovered after detrending.
We show the errors for each photometric measurement before detrending, 
propagated from the errors estimated  by the TAOS photometry pipeline \citep{Zhang2009}.

\subsection{Detection of  Short-Period Variable Stars}

After we finished the preprocessing, explained in the previous section, 
we applied the Fast Fourier Transform  algorithm (FFT; \citealt{Brigham1974}) to each light-curve in order to detect periodic signals.
Note that the individual measurements of TAOS light-curves are evenly spaced with a 5Hz sampling rate.\footnote{Binned light-curves are evenly spaced as well.}
Thus FFT is appropriate for the detection of periodic signals.
We focused on the detection of short-period variable stars whose periods are $\simless 1.5$ hour  because TAOS monitors a given field for a maximum  of 1.5 hours.

We describe  the basic steps of the detection process below:

\begin{itemize}

\item We apply FFT to each detrended light-curve and derive the power spectrum of the light-curve. 
\label{item:FFT1}
We then examine if there exists a frequency (or frequencies) whose power is bigger than five times the standard deviation of powers of the background frequencies.
The standard deviation of powers is calculated  after removing outliers using 3-sigma clipping.\footnote{Those outliers are only removed for the calculation of the standard deviation. They are included in the search of periodic signals.}
We identify the star as a variable candidate if there is a  frequency 
higher than five times the standard deviation. 
\label{item:FFT2}

\item For each candidate variable we  check  if the periodic signal is detected on the other telescopes' light-curves of the same run.
If it is not detected by the other telescopes, we remove the star from the candidate list.
\label{item:FFT5}

\item We visually inspect all  raw zipper images for the candidates and remove false positives
caused by moving asteroids, photometry defects or other contamination due to various noise sources.
For instance, the flux of stars in the neighborhood of fast moving objects could 
be increased and decreased within an hour, which resembles periodic signals.
\label{item:FFT6}

\item  We cross-match all of the candidates with SIMBAD \citep{Wenger2000}
and remove the false positives that are confirmed to be other types of variable stars (e.g. eclipsing binary stars).

\item Finally we  remove the variable stars whose periods are longer than 1.5 hours.

\end{itemize}

\section{Detection Results}
\label{sec:DetectionResults}

With the detection algorithm described in the previous section, we found
41 $\delta$ Sct candidate stars  whose periods are shorter than 1.5 hours and whose amplitudes are within a few hundredth of a magnitude (hereinafter, TAOS $\delta$ Sct stars).
Among those 41 TAOS $\delta$ Sct stars, one of them is a previously {\em suspected} variable star, NSV 3816, from the Suspected Variable stars and Supplement \citep{Samus2009}. However, the period and amplitude of NSV 3816 have never been published before.
The remaining 40 TAOS $\delta$ Sct stars are newly detected by this study.

After the TAOS $\delta$ Sct stars were identified, we extracted the physical parameters of each
star by cross-matching them with various astronomical catalogs.
We show catalogs we used in Table \ref{tab:Catalogs_Parameters}.
We found that 12 of the detected 41 TAOS $\delta$ Sct stars have
spectral types from A0 to F5, which are typical spectral types for $\delta$ Sct stars.
Unfortunately the rest of them, except for two peculiar $\delta$ Sct stars -discuss in Section \ref{sec:Spectroscopy}-,
do not have spectral information.
Nevertheless, their short period and low amplitude strongly suggest
that they are LADS rather than other types of variables, such as RR Lyrae or Cepheids, whose periods and amplitudes are relatively longer and larger than those of $\delta$ Sct stars.

\begin{table*}[t]
 \begin{center}
 \caption{Catalogs Used to Extract Additional Parameters \label{tab:Catalogs_Parameters}}
 \begin{tabular}{cc}
\tableline\tableline
 Catalog & Reference \\
\tableline
 
GCVS & \citet{Perryman1997} \\
All-Sky Compiled Catalogue of 2.5 million stars & \citet{Kharchenko2001} \\
HD &  \citet{Cannon1993} \\
Catalogue of Stellar Spectral Classifications & \citet{Skiff2009} \\
Tycho-2 Catalogue of the 2.5 Million Brightest Stars &\citet{Hog2000} \\
Guide Star Catalog (GSC) & \citet{Lasker2008} \\
USNO-B 1.0 & \citet{Monet2003} \\
SAO Star Catalog J2000 & \cite{SAO1995} \\
Catalog of Projected Rotational Velocities & \citet{Glebocki2000} \\
Rotational Velocity Determinations for 118 $\delta$ Sct Variables & \citet{Bush2008} \\

  \tableline
 \end{tabular}
 \end{center}
 \end{table*}

As a byproduct of our analysis, we detected a previously known variable star with  $\delta$ Sct  pulsation, GM Leo, 
which is actually a $\lambda$ Bootis star \citep{Handler2000}.
Some  $\lambda$ Bootis stars show $\delta$ Sct  pulsations \citep{Paunzen2004}
and can have spectral types from late B to early F, 
which makes it difficult to distinguish them from $\delta$ Sct stars.
In such cases,  there are no clear differences between $\delta$ Sct stars
and $\lambda$ Bootis stars except the metal abundance \citep{Balona2004};
$\lambda$ Bootis stars show weak metal lines such as Mg {\scriptsize{II}} $\lambda$4481 line \citep{Paunzen2004}.

We also checked several preexisting catalogs of $\delta$ Sct stars to see
if there are previously known $\delta$ Sct stars in the TAOS observation fields.
Table \ref{tab:Catalogs} shows the preexisting catalogs we checked.
As a result of this search, we found only one previously known 
$\delta$ Sct  pulsation star to be in the TAOS observation fields.
That turns out to be GM Leo, which as we mentioned above, we successfully detected.
Although GCVS classified GM Leo as $\delta$ Sct star based on the work by \citet{Handler2000}, 
\citet{Handler2000} in their paper claimed GM Leo is not a $\delta$ Sct star but a $\lambda$ Bootis star.
Thus we removed GM Leo from our detection list.

 \begin{table*}[t]
 \begin{center}
 \caption{Preexisting Catalogs of $\delta$ Sct Stars  \label{tab:Catalogs}}
 \begin{tabular}{cccc}
\tableline\tableline
 Catalog & Source Surveys & Number of $\delta$ Sct Stars & Reference \\
\tableline
 
R2000$^a$ & MACHO, OGLE, Hipparcos, etc & $\sim$600 & \citep{Rodriguez2000} \\
ROTSE$^b$ & ROTSE & 6 & \citep{Jin2003}\\
ASAS$^c$ & ASAS & $\sim$500 & \citep{Pojmanski2006} \\
GCVS$^d$ & various surveys & $\sim$500 & \citep{Samus2009}\\
TAOS & TAOS & 41 & this paper \\

& & & \\

Others & & & \\
& 5 new $\gamma$ Doradus and 5 new $\delta$ Sct survey & 5 & \citep{Henry2001} \\
& Case study for HD 173977 & 1 & \citep{Chapellier2004} \\
& Case study for HD 8801  & 1 & \citep{Henry2005} \\
& The first HADS in an eclipsing binary star & 1 & \citep{Christiansen2007} \\
& Variable stars in NGC 2099 & 9 & \citep{Kang2007} \\
& Transit survey of M37 & 2 & \citep{Hartman2008} \\
& ASAS variable stars in the Kepler field of view & 4 & \citep{Pigulski2009} \\

  \tableline
 \end{tabular}
 \end{center}
  \begin{flushleft}
  $^a$ The catalog compiled by \citet{Rodriguez2000}\\ 
  $^b$ Robotic Optical Transient Search Experiment. \\
  $^c$ All-Sky Automated Survey. \\
  $^d$ General Catalog of the Variable Stars.
   \end{flushleft}
 \end{table*}

\subsection{List of the Detected 41 $\delta$ Sct Stars}
\label{sec:ListDeltaScuti}

Table \ref{tab:Delta_Scuti} shows the 41 TAOS $\delta$ Sct stars' physical information such as positions,
magnitudes, frequencies, amplitudes, spectral types, etc.

\begin{figure}[tbp!]
\begin{center}
       \includegraphics[width=0.45\textwidth]{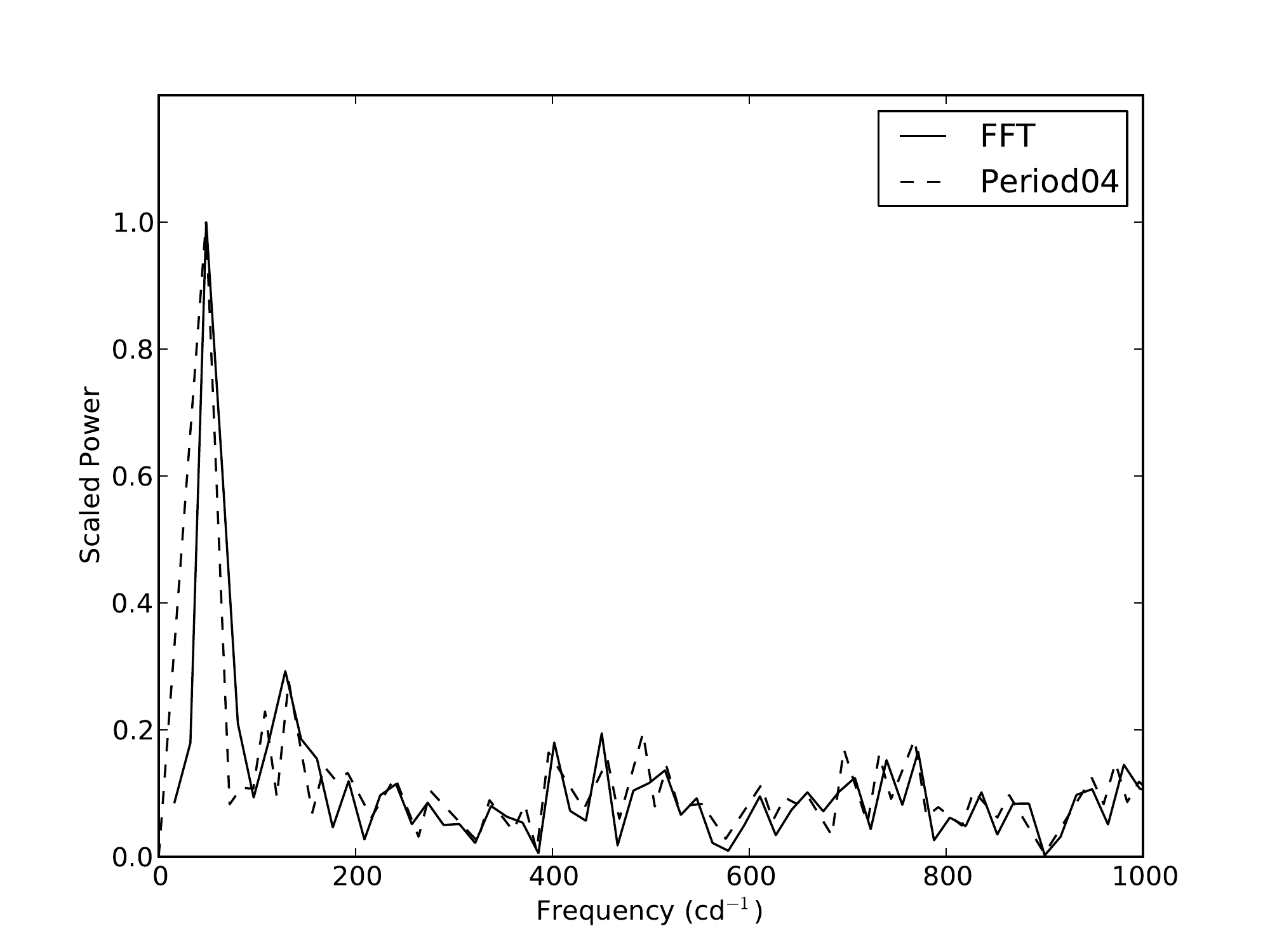}
\end{center}
    \caption{A comparison result of power spectrum derived from an FFT method and from {\fontfamily{pcr}\selectfont PERIOD04}.
X-axis is frequency in counts/day, y-axis is scaled power. 
Solid line is the power spectrum derived from an FFT method and dashed line is the power spectrum derived from {\fontfamily{pcr}\selectfont PERIOD04}. The two spectra appear almost identical.}
    \label{fig:CompareSpectrum}
\end{figure}

We used the FFT algorithm to detect periodic signals however we used {\fontfamily{pcr}\selectfont PERIOD04} to derive their physical parameters
such as period and amplitude.\footnote{Since {\fontfamily{pcr}\selectfont PERIOD04} gives half of the full amplitudes, we doubled amplitudes derived by {\fontfamily{pcr}\selectfont PERIOD04} as \citet{Rodriguez2000} and other authors do.} 
 This is because {\fontfamily{pcr}\selectfont PERIOD04} improves the frequency by fitting the  light-curve with a combination of sine curves. 
Moreover {\fontfamily{pcr}\selectfont PERIOD04}   also provides errors for the derived frequencies. 
Figure \ref{fig:CompareSpectrum} shows a comparison result of power spectra derived from an FFT method and {\fontfamily{pcr}\selectfont PERIOD04}  for a single TAOS $\delta$ Sct star.
X-axis is frequency in counts/day, y-axis is scaled power. Solid line is power spectrum derived from an FFT method and dashed line is power spectrum derived from {\fontfamily{pcr}\selectfont PERIOD04}. 
 As the figure shows, the two spectra are almost consistent.

As we mentioned in the previous section, TAOS operates multiple telescopes simultaneously monitoring the same patch of the sky.
Thus we have a maximum of  three simultaneous light-curves for all TAOS $\delta$ Sct stars for a given zipper run.
To derive more precise frequencies and amplitudes, for each identified $\delta$ Sct star we summed the light-curves from each of the telescopes.
Moreover, the TAOS telescopes occasionally visit same fields  multiple times.
In such cases, we merge all corresponding normalized light-curves\footnote{We normalized each light-curve by their mean values.}  
of each identified $\delta$ Sct star into a single but longer light-curve.
Having longer light-curves we were able to extract multiple frequencies using {\fontfamily{pcr}\selectfont PERIOD04}. 
Among the extracted frequencies, we selected the frequencies whose SNR is bigger than  five.
The SNR of each frequency was calculated using {\fontfamily{pcr}\selectfont PERIOD04} as well.\footnote{Although other authors have suggested a threshold of SNR $>4$  \citep{Breger1993, Christiansen2007}, we empirically found that a threshold of SNR $>$ 4 produces  false positives and thus we set the SNR threshold to five.}
 As a result we found 16 TAOS $\delta$ Sct stars having multiple frequencies (see Table \ref{tab:Delta_Scuti}).
Note that we did not attempt to extract multiple frequencies if the star is detected only once (i.e. detected in only single zipper run).
It is also worth mentioning that we lose detectability on relatively long period pulsations because we normalized the light-curves while merging them.

All detected $\delta$ Sct stars are relatively bright as shown in the table (the faintest star's $m_V$ is around 12).
This is because the limiting magnitude of TAOS {\em zipper mode} is relatively bright at  $\sim$13.5 \citep{Lehner2009}
and also because high SNR is needed to detect low-amplitude
variations of a few milli-magnitude.
To find $m_V$, $m_B$ and spectral types, we used
Centre de Donne'es astronomiques de Strasbourg (CDS) web service \citep{Genova2000}.

In Table \ref{tab:Delta_Scuti}, we also provide the number of the total observations (i.e. the number of the total zipper runs) and the number of identifications by the FFT analysis for each TAOS $\delta$ Sct star.
Since the TAOS telescopes observe same fields multiple times, 
we have multiple light-curves for stars in the fields. 
Note that we applied the FFT algorithm to each light-curve to detect periodic signals.
Nevertheless, as the table shows, not every light-curve of the TAOS $\delta$ Sct stars were confirmed to have periodic signals.
This is mainly because of the poor quality of some  of the light-curves caused by trends and noise (i.e. unstable weather, telescope vibration, etc).
Although we removed most of the trends using PDT, it is nearly impossible to recover the intrinsic periodic signal of a few milli-magnitude in the presence of large systematic errors.

\subsection{Spectral Windows and Power Spectra of TAOS $\delta$ Sct Stars}

\begin{figure}[tbp!]
\begin{center}
       \includegraphics[width=0.45\textwidth]{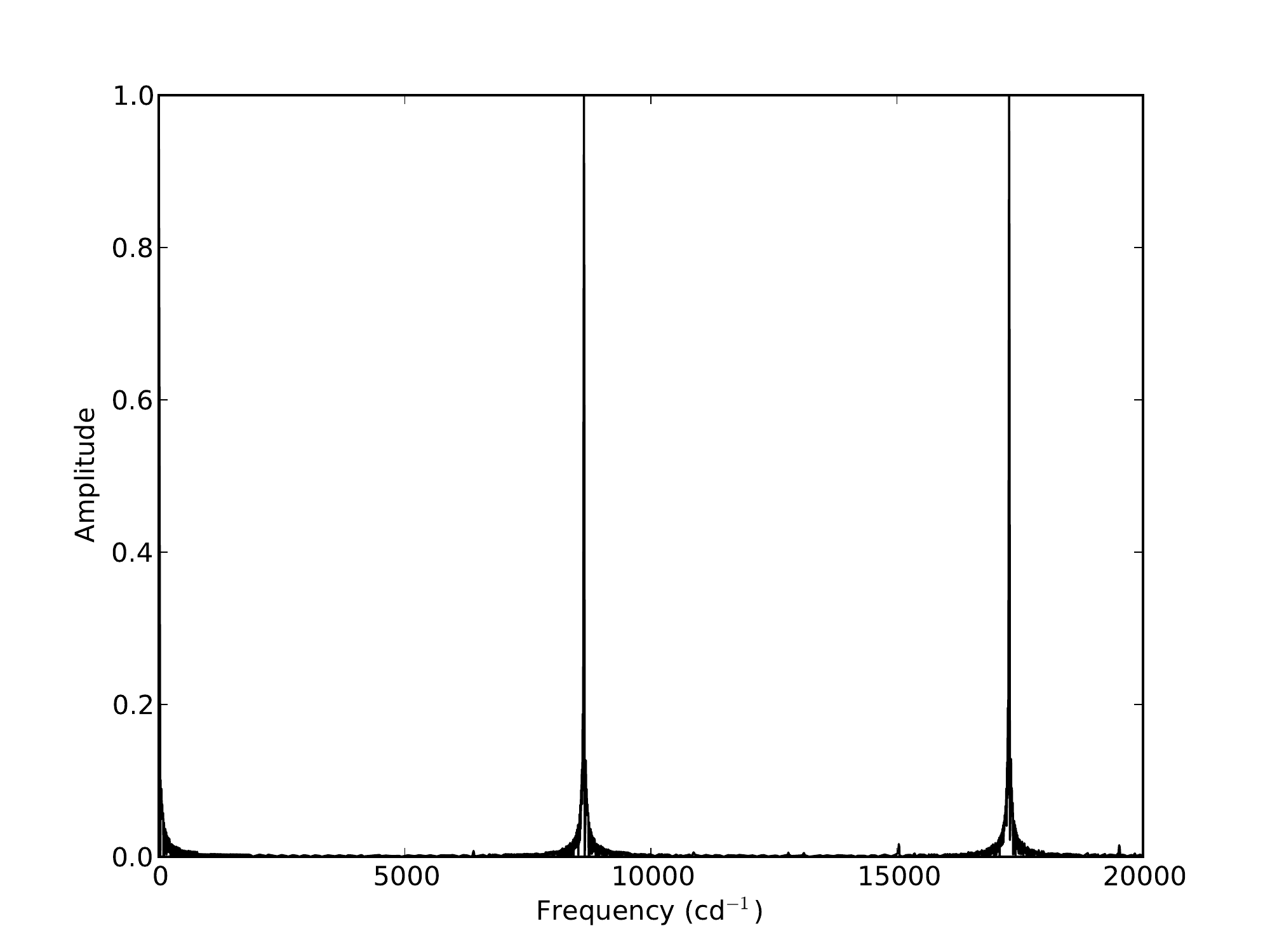}
\end{center}
    \caption{An example of a spectral window of a single zipper run. Peaks with regular intervals (two times of the Nyquist frequency) appear
    because points are equally spaced  in a single zipper run.}
    \label{fig:Nyquist_sw}
\end{figure}

 In Figure \ref{fig:Nyquist_sw} we show an example of a spectral window of a single zipper run. 
Since the observational times  are almost equally spaced, 
peaks with regular intervals (two times the Nyquist frequency)  are present in the spectral window \citep{Deeming1975}.
The Nyquist frequency is 4320Hz since the gap between each consecutive binned data is 10 sec.

\begin{figure}[t]
\begin{center}
       \includegraphics[width=0.38\textwidth,angle=90]{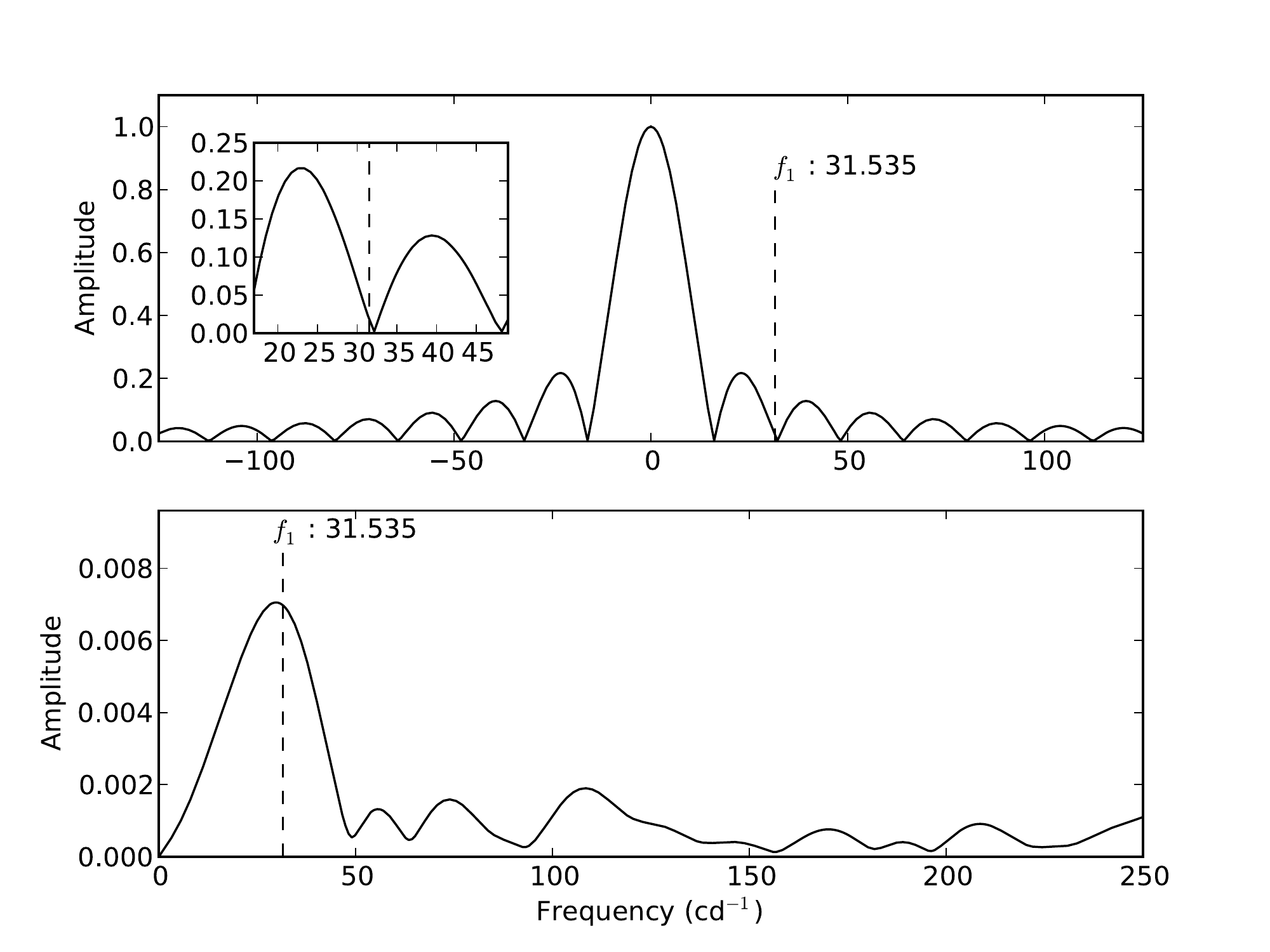}
\end{center}
    \caption{The spectral window and the power spectrum of the star ID 020.00141. 
    The top panel shows the spectral window  and the bottom panel shows the power spectrum of the star. 
    The dashed line shows the detected frequency.
    In the top panel, we magnified the spectral window to clearly show the detected frequency.}
    \label{fig:Worst_PS}
\end{figure}

\begin{figure}[t]
\begin{center}
       \includegraphics[width=0.38\textwidth,angle=90]{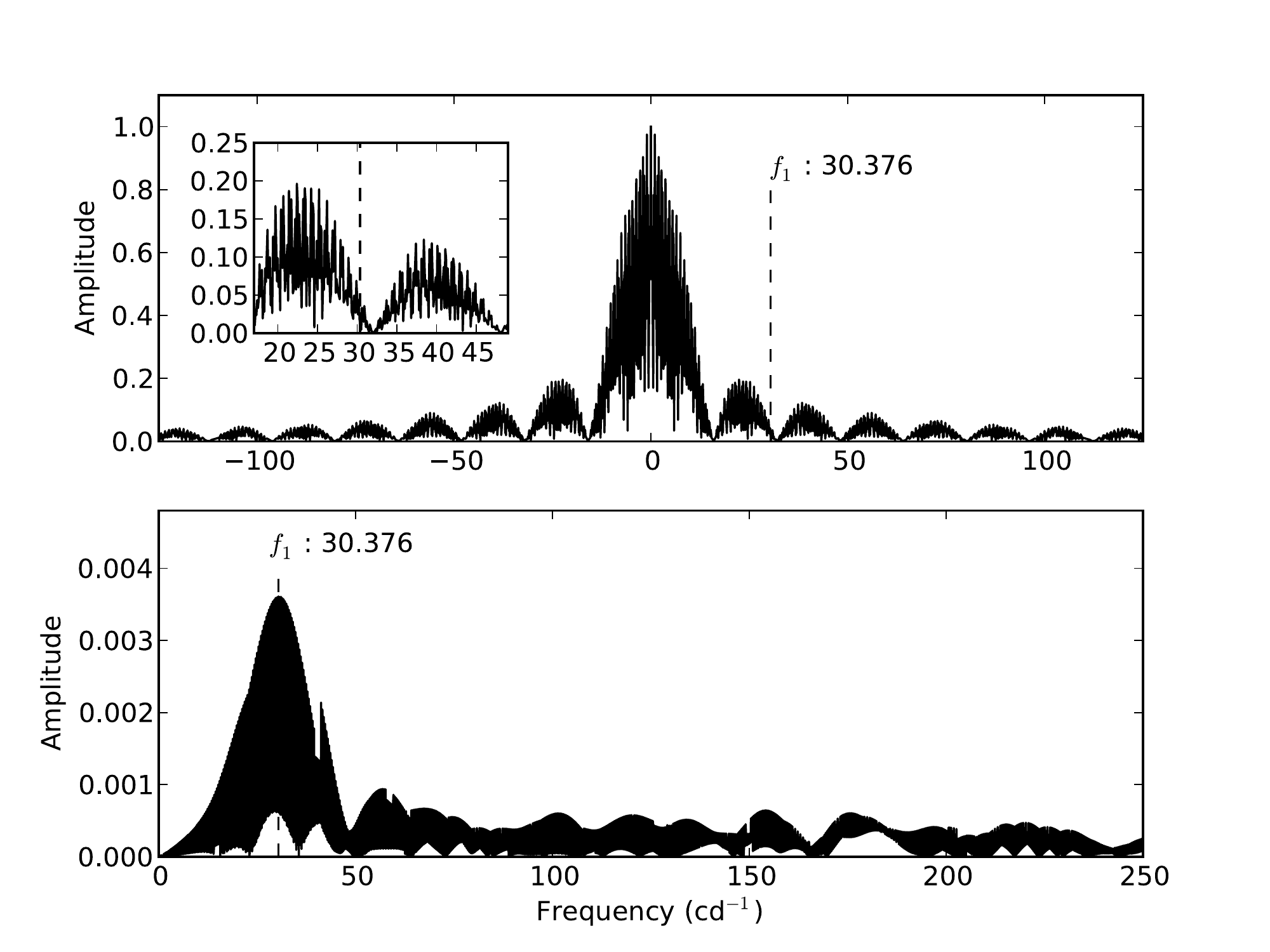}
\end{center}
    \caption{The spectral window and the power spectrum of the star ID 121.00043.
    The top panel shows the spectral window  and the bottom panel shows the power spectrum of the star. 
    The dashed line shows the detected frequency.
    In the top panel, we magnified the spectral window to clearly show the detected frequency.}
    \label{fig:Inter_PS}
\end{figure}

\begin{figure}[tp!]
\begin{center}
       \includegraphics[width=0.5\textwidth]{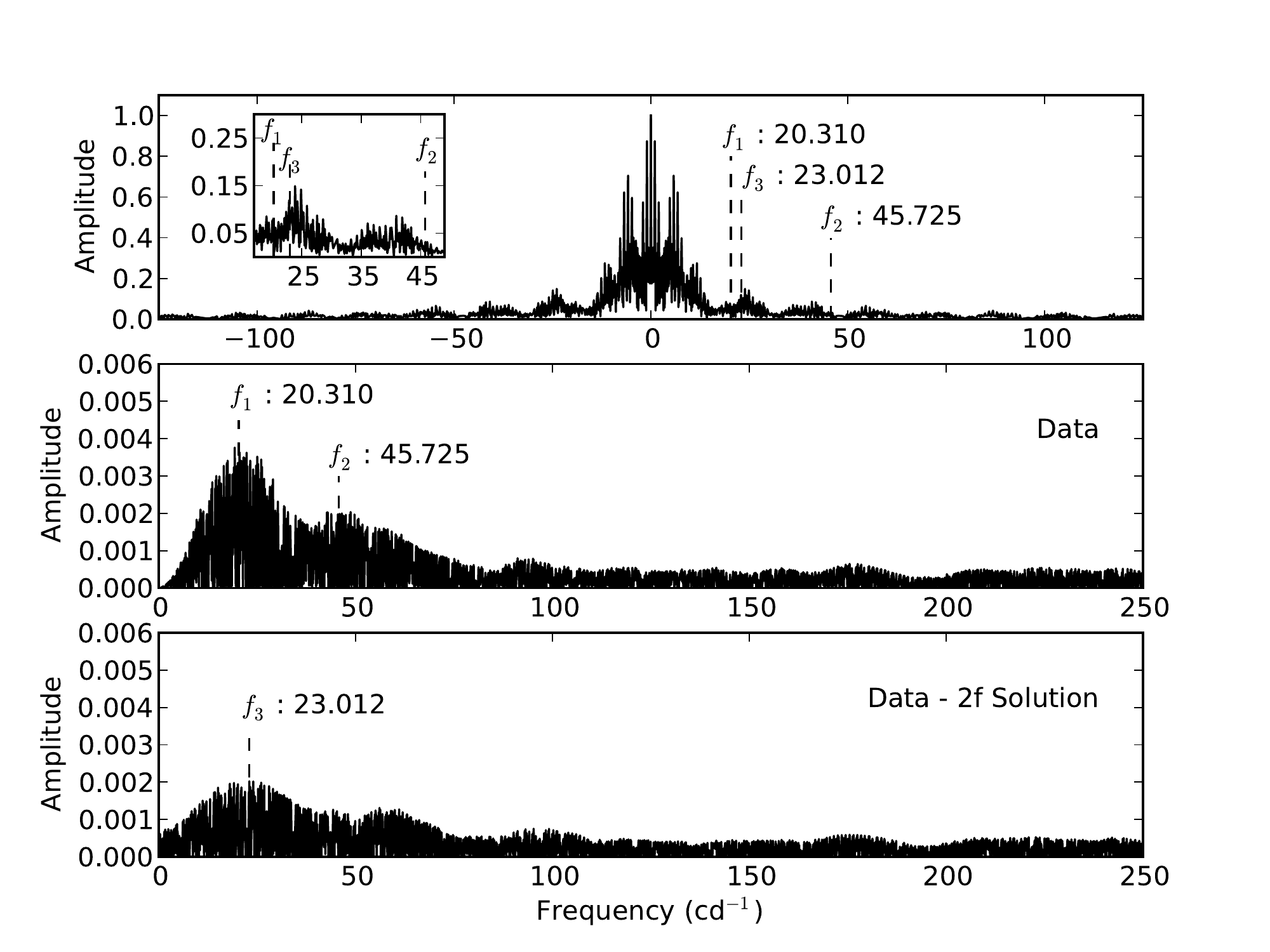}
\end{center}
    \caption
           {The spectral window and the power spectrum of the star ID 054.00014. 
           We detected three frequencies (dashed lines) using {\fontfamily{pcr}\selectfont PERIOD04}.
	  The top panel shows the spectral window.
	  The middle panel shows the first two frequencies.
           The bottom panel is the power spectrum after whitening the two frequencies.
           In the top panel, we magnified the spectral window to clearly show the detected frequencies.
}
    \label{fig:Best_PS}
\end{figure}

Figure \ref{fig:Worst_PS}, \ref{fig:Inter_PS} and \ref{fig:Best_PS} show the spectral windows along with the power spectra of three TAOS $\delta$ Sct stars.
Stars  included in the figures are 020.00141 (Figure \ref{fig:Worst_PS}), 121.00043 (Figure \ref{fig:Inter_PS}) and 054.00014 (Figure \ref{fig:Best_PS}). 
The top panels in each figure show the spectral windows and the bottom panels (and the middle panel in Figure \ref{fig:Best_PS}) show the power spectra of the stars. Dashed lines indicate the detected frequencies. 
Note that we improved the detected frequencies by fitting a combination of sine waves using {\fontfamily{pcr}\selectfont PERIOD04}.
Thus the improved frequencies  could be slightly shifted from the original peaks  in the  power spectra (e.g. see the bottom left panel in Figure \ref{fig:Worst_PS}) after the fitting.

As Table \ref{tab:Delta_Scuti} shows, the star with ID 020.00141 was identified only once and is relatively fainter  ($m_V=$ 11.40) than other TAOS $\delta$ Sct stars. Its amplitude  is one of the smallest amplitudes and  its detected frequency SNR  is the lowest among TAOS $\delta$ Sct stars. Thus the power spectrum of this star represents one of the ``worst case scenario''. We detected one single frequency for this star. 
The star with ID 121.00043 was identified three times and is relatively bright ($m_V$ = 9.13).  The spectral window and the power spectrum of the star could represent a ``moderate-level scenario''. We detected one frequency for this star.
Finally, the star with ID 054.00014 was identified nine times and is relatively bright ($m_V$ = 9.50). Thus its spectral window and power spectrum represents one of the ``best case scenario''. Using {\fontfamily{pcr}\selectfont PERIOD04}, we detected three frequencies for this star.  
As  can be seen from the figures, there are no significant peaks in the spectral windows at the detected frequencies.

\subsection{Spectroscopy of  Two Peculiar Spectral Type $\delta$ Sct Stars}
\label{sec:Spectroscopy}

As we mentioned in the previous section, we found two peculiar spectral type $\delta$ Sct stars
which have B8 and G5 spectral types. 
These are the bluest and reddest spectral types of $\delta$ Sct stars ever detected.
The spectral type of the B8 star was extracted from the Henry Draper Catalogue and Extension (HD, \citealt{Cannon1993}).
Unfortunately we could not find any spectroscopic literature for the G5 star so we suspect that the spectral type is most likely
derived  from its color information. The spectral type of the G5 star was extracted from SIMBAD.
To confirm their spectral types, we observed the two stars with spectroscopic instruments.

\begin{figure*}[t]
\begin{center}
       \includegraphics[width=1\textwidth]{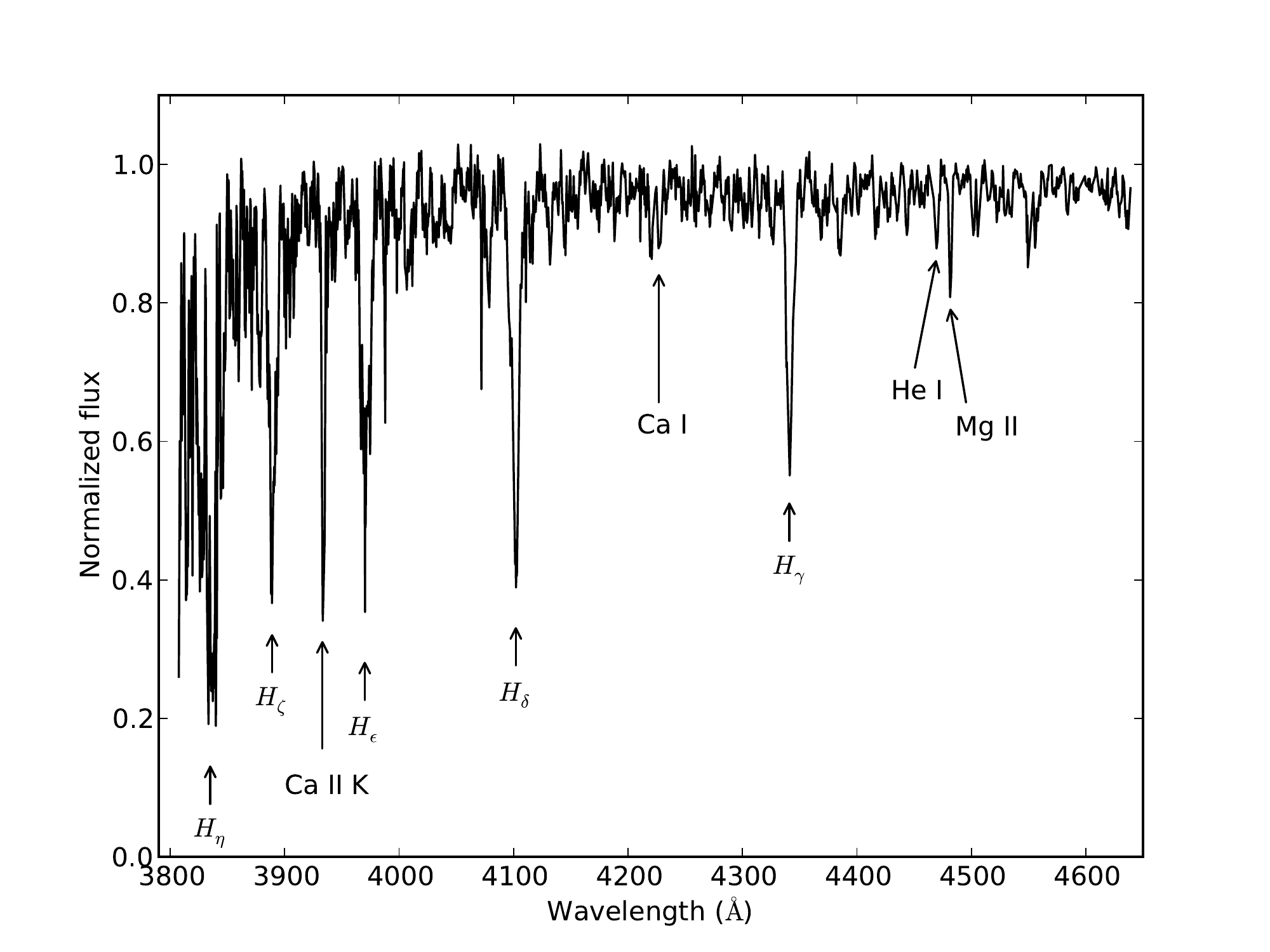}
\end{center}
    \caption{The normalized spectrum of the B8 candidate star. There are strong Ca II K line and weak metallic lines, which is typical for A type stars.
    The star is likely an A5 type star rather than a B8 type star.}
    \label{fig:spectrum}
\end{figure*}

For the B8 star, we used the BOES\footnote{BoaO Echelle Spectrograph}
of the 1.8-m telescope at the Bohyunsan Optical Astronomy Observatory (BOAO), South Korea \citep{Kim2007}.
We used {\fontfamily{pcr}\selectfont IRAF} \citep{Tody1986, Tody1993}
for the reduction of the obtained spectroscopic data.
Figure \ref{fig:spectrum} shows the normalized spectrum of the B8 candidate star.
We indicate several important spectral lines in the figure.
As the figure shows, Ca {\scriptsize{II}} K line is very strong which is typical for A type stars \citep{Gray1987}.
B type stars do not show such strong Ca {\scriptsize{II}} K line.
The spectrum also shows weak metallic lines (e.g. Ca {\scriptsize{I}} and Mg {\scriptsize{II}} line)
which are usually presented in A type stars.
Based on the  strength of Ca {\scriptsize{II}} K line, hydrogen lines and metallic lines, the star is likely 
an A5 type star although the classification of sub-class is rather uncertain due to the low SNR of the spectrum.

In addition, to observe the G5 star, we used the FAST instrument mounted at the Fred Lawrence Whipple Observatory (FLWO)
1.5m telescope, Mount Hopkins in Arizona \citep{Fabricant1998}.
After comparing the observed data with standard spectral libraries \citep{Pickles1998},
we found that the spectral type of the star is not a G5 but an F0.
Therefore the star is likely a typical $\delta$ Sct star.

\section{Summary}
\label{sec:Conclusion}

We analyzed the TAOS 2-year data accumulated during 2005 and 2006 observations
in order to find short-period variables. Using the TAOS photometry pipeline we created photometric light-curves.
We removed systematic trends commonly appeared in the light-curves using PDT.
To detect periodic signals in the detrended light-curves, we applied the Fast Fourier Transform (FFT) to each light-curve.
FFT is a simple but {\em powerful} algorithm for detection of periodic signals when data points are evenly spaced.
We then chose light-curves which possess a frequency (or frequencies) whose power is five times larger than the standard deviation of  powers of all background frequencies in the power spectrum derived using FFT.
We visually checked the light-curves and raw images of all candidates to remove false positives caused by moving asteroids, photometry defects, etc.
We also removed candidates which were detected by only one of the three telescopes.
All  remaining 41 variable candidates have periods about an hour and amplitudes less than a few hundredth of a magnitude,
which strongly suggests that they are low-amplitude $\delta$ Sct stars (LADS).

We cross-matched the detected $\delta$ Sct candidate stars with many astronomical
catalogs to extract additional information
(e.g. magnitude, spectral type, variability type, etc).
As a result we found that 14 stars have spectral types from A to F,
which are typical spectral types for $\delta$ Sct stars.
The rest of the detected $\delta$ Sct stars do not have spectral information.

The light-curves of TAOS $\delta$ Sct stars
are  accessible at the Time Series Center (TSC, http://timemachine.iic.harvard.edu),
Initiative in Innovative Computing (IIC) at Harvard.
{\fontfamily{pcr}\selectfont PERIOD04} project files of each star is also provided.
The project files contain complete light-curve data, power spectrum, frequency and amplitude information.

\section*{Acknowledgements}

Y.-I. Byun acknowledges the support of National Research Foundation of Korea through Grant 2009-0075376.
The work at National Central University was supported by the grant NSC 96-2112-M-008-024-MY3.
Work at Academia Sinica was supported in part by the thematic research program AS-88-TP-A02. Work at the Harvard College
Observatory was supported in part by the National Science Foundation under grant AST-0501681 and by NASA under grant NNG04G113G. SLM's work was performed under the auspices of the U.S. Department of Energy by Lawrence Livermore National Laboratory in part under Contract
W-7405-Eng-48 and by Stanford Linear Accelerator Center under Contract DE-AC02-76SF00515.  K. H. Cook's work was performed under the auspices of the U.S. Department of Energy by Lawrence Livermore National Laboratory in part under Contract W-7405-Eng-48 and in part under Contract DE-AC52-07NA27344.
We also thanks J. D. Hartman at Harvard-Smithsonian Center for Astrophysics for useful discussion.

The detrending and the analysis of datasets in this paper were run on the
Odyssey cluster supported by the FAS Research Computing Group at the Harvard.
This research has made use of the SIMBAD database, operated at CDS, Strasbourg, France.
IRAF is distributed by the National Optical Astronomy Observatories, which are operated by the Association of Universities for Research    in Astronomy, Inc., under cooperative agreement with the National Science Foundation.

\vspace{0.5cm}
{\small{
\bibliography{TAOS_Delta_Scuti}{}
}}

\onecolumn

\begin{landscape}
\renewcommand{\arraystretch}{1.1}
{\small {
\begin{center}
\begin{longtable}{ccccccccccccc}
\caption{TAOS $\delta$ Sct stars in TAOS 2-year Data } \label{tab:Delta_Scuti} \\

\tableline\tableline
\multicolumn{1}{c}{No} &
\multicolumn{1}{c}{ID$^a$} &
\multicolumn{1}{c}{RA} &
\multicolumn{1}{c}{Dec} &
\multicolumn{1}{c}{$m_V$} &
\multicolumn{1}{c}{$m_B$} &
\multicolumn{1}{c}{Frequency} &
\multicolumn{1}{c}{$\Delta m_V^b$} &
\multicolumn{1}{c}{SNR$^c$} &
\multicolumn{1}{c}{Epoch} &
\multicolumn{1}{c}{Spectral Type} &
\multicolumn{1}{c}{\#/\#$^d$} &
\multicolumn{1}{c}{Note } \\
&  & \multicolumn{1}{c}{(hh:mm:ss)} & \multicolumn{1}{c}{(dd:mm:ss)} & &   &\multicolumn{1}{c}{(cd$^{-1}$)} &\multicolumn{1}{c}{(mmag)} & & \multicolumn{1}{c}{(MJD)} & & \\
\tableline
\endfirsthead

\multicolumn{11}{l}
{{\bfseries \tablename\ \thetable{} -- continued from previous page}} \\
\tableline\tableline
\multicolumn{1}{c}{No} &
\multicolumn{1}{c}{ID$^a$} &
\multicolumn{1}{c}{RA} &
\multicolumn{1}{c}{Dec} &
\multicolumn{1}{c}{$m_V$} &
\multicolumn{1}{c}{$m_B$} &
\multicolumn{1}{c}{Frequency} &
\multicolumn{1}{c}{$\Delta m_V^b$} &
\multicolumn{1}{c}{SNR$^c$} &
\multicolumn{1}{c}{Epoch} &
\multicolumn{1}{c}{Spectral Type} &
\multicolumn{1}{c}{\#/\#$^d$} &
\multicolumn{1}{c}{Note } \\
&  & \multicolumn{1}{c}{(hh:mm:ss)} & \multicolumn{1}{c}{(dd:mm:ss)} & &   &\multicolumn{1}{c}{(cd$^{-1}$)}  &\multicolumn{1}{c}{(mmag)} & &\multicolumn{1}{c}{(MJD)}  & & \\
\tableline\endhead

\tableline \multicolumn{13}{r}{{Continued on next page}}
\endfoot

\tableline
\endlastfoot

1 & 124.00003 & 00:52:40 & +06:39:55 &  8.89 &  9.20 &24.421 $\pm$  3.975e$^{-4}$ & 3.83 $\pm$ 0.20 & 6.5 &  53626.7469  &           A2 &4 / 7&\\

2 & 038.00124 & 02:56:54 & +34:23:20 & 11.60 & 12.00 &23.598 $\pm$ 4.775e$^{-1}$ & 7.95 $\pm$ 0.44 & 11.9 & 53678.6909    &              &1 / 2&\\

3 & 053.00009 & 03:37:02 & +18:21:51 &  8.48 &  8.80 & 20.544 $\pm$ 2.314e$^{-4}$ & 5.99 $\pm$  0.14 & 18.5 &     53671.7396  &         A2 &3 / 8&\\

4 & 059.00115 & 03:42:41 & +17:55:01 & 12.12 & 12.60 & 25.836 $\pm$ 7.587e$^{-4}$ &  9.19 $\pm$ 0.68 &  13.6    &  54021.6614   &        &3 / 8&\\
  &  &  & &  & & 32.116 $\pm$ 8.434e$^{-4}$ & 7.97 $\pm$  0.66 & 11.8    &    54021.6579         &  & &\\
  &  &  & &  & &12.034 $\pm$ 1.291e$^{-3}$ & 4.94 $\pm$  0.58 & 7.5    &    54021.5979         &  & &\\
    &  &  & &  & &40.403 $\pm$ 1.713e$^{-3}$ & 3.53 $\pm$  0.62 & 5.25    &    54021.6523         &  & &\\

5 & 059.00005 & 03:46:01 & +18:34:00 &  9.14 &  9.46 & 33.115 $\pm$ 2.022e$^{-4}$ & 8.00 $\pm$  0.20 &  15.7    &  54006.7013   & A5       &7 / 8&B8$^e$\\
  &  &  & &  & & 18.068 $\pm$ 4.793e$^{-4}$ & 3.26 $\pm$  0.20 & 6.4    &  54006.7020         &  & &\\

6 & 049.00056 & 04:03:21 & +19:21:31 & 10.91 &  11.48 &  24.205 $\pm$ 3.694e$^{-4}$ & 9.15 $\pm$ 0.34 &13.7 & 54029.8147   &               &6 / 16&\\
  &  &  & &  & & 17.761 $\pm$ 6.312e$^{-4}$ & 4.98 $\pm$  0.36 & 7.5    &  54029.7806         &  & &\\
    &  &  & &  & & 29.544 $\pm$ 8.243e$^{-4}$ & 4.00 $\pm$  0.34 & 6.0    &  54029.8094         &  & &\\

7 & 068.00053 & 04:30:06 & +20:55:00 & 11.45 & 12.36 & 32.504 $\pm$ 2.402e$^{-5}$ & 10.17 $\pm$ 0.28 & 15.5 &  53643.8513    &           F0 &8 / 15&G5$^f$\\
    &  &  & &  & & 26.117 $\pm$ 5.939e$^{-5}$ & 4.10 $\pm$  0.28 & 6.24    &  53643.8542         &  & &\\

8 & 060.00151 & 04:48:37 & +21:10:33 & 11.44 & 11.64 & 43.180 $\pm$ 3.709e$^{-4}$ & 5.13 $\pm$ 0.26 & 8.4 & 54012.7340     &          F5 &6 / 32&\\

9 & 022.00001 & 04:56:14 & +21:34:20 &  7.34 &  7.70 & 21.582 $\pm$ 1.709e$^{-4}$ & 3.97 $\pm$ 0.12 & 12.9 & 54021.7647       &        F0 &7 / 25&\\
 &  &  & &  & & 21.146 $\pm$ 3.268e$^{-4}$ & 2.15 $\pm$  0.12 & 6.94    &  54021.7495         &  & &\\
  &  &  & &  & & 40.637 $\pm$ 4.339e$^{-4}$ & 1.57 $\pm$  0.12 & 5.12    &  54021.7926         &  & &\\

10 & 020.00206 & 05:08:38 & +22:49:37 & 11.85 & 12.16 & 53.575 $\pm$ 5.075e$^{-1}$ & 7.62 $\pm$ 0.46 & 9.9 & 53705.6765    &              &1 / 3&\\

11 & 020.00141 & 05:09:24 & +23:16:05 & 11.40 & 12.29 & 31.535 $\pm$ 6.862e$^{-1}$ & 6.87 $\pm$ 0.52 & 5.7 & 53705.6877      &            &1 / 3&\\

12 & 021.00011 & 05:09:40 & +21:50:05 &  8.99 & 9.21 & 41.754 $\pm$ 1.493e$^{-4}$  & 8.05 $\pm$ 0.18  & 17.7 &  53975.7771        &         &5 / 7&\\
  &  &  & &  & & 37.264 $\pm$ 4.476e$^{-4}$ & 2.73 $\pm$  0.18 & 5.9    &  53975.7883         &  & &\\

13 & 020.00135 & 05:10:14 & +23:01:24 & 10.90 & 11.78  & 39.254 $\pm$ 6.865e$^{-1}$ & 5.11 $\pm$ 0.40 & 6.3 & 53705.6974     &             &1 / 3&\\

14 & 024.00234 & 05:15:19 & +22:53:41 & 11.93 & 12.39  & 25.804 $\pm$ 2.471e$^{-1}$ &  35.35 $\pm$ 1.04 & 12.8 &  53679.7864     &            &1 / 2&\\

15 & 160.00106 & 06:01:30 & +21:27:38 & 10.38 & 10.69  & 21.063 $\pm$ 3.878e$^{-4}$ &  7.44 $\pm$ 0.22 & 13.9 & 53680.8125      &            &3 / 8&\\

16 & 160.00004 & 06:04:05 & +21:29:39 &  7.81 &  7.97 & 25.827 $\pm$ 5.607e$^{-4}$ & 4.70 $\pm$ 0.16 & 15.5 &  53680.8193        &         &3 / 8&\\
  &  &  & &  & & 22.395 $\pm$ 7.432e$^{-4}$ & 3.55 $\pm$  0.14 & 11.7    &  53680.8406         &  & &\\
    &  &  & &  & & 38.868 $\pm$ 1.077e$^{-3}$ & 1.59 $\pm$  0.12 & 5.2    &  53680.8298         &  & &\\
    
17 & 160.00199 & 06:04:26 & +21:21:55 & 10.96 & 11.16 & 43.741 $\pm$ 9.070e$^{-4}$ &  4.12 $\pm$ 0.26 & 11.0 & 53680.8119      &            &3 / 8&\\

18 & 052.00132 & 07:39:07 & +21:39:20 & 11.14 & 11.67 & 23.802 $\pm$ 3.343e$^{-3}$  & 4.69 $\pm$ 0.30 & 7.2 &  53699.7766       &          &2 / 3&\\

19 & 052.00069 & 07:39:09 & +20:48:41 & 10.43 & 10.85 & 21.288 $\pm$ 7.664e$^{-4}$&  7.51 $\pm$  0.28 & 11.6 &  53682.7645      &           &3 / 3&\\
    &  &  & &  & & 29.627 $\pm$ 1.666e$^{-3}$ &  3.41 $\pm$  0.28 & 5.2    &  53682.7835         &  & &\\

20 & 052.00159 & 07:39:20 & +21:11:22 & 11.25 & 11.59 & 18.793 $\pm$ 2.293e$^{-3}$ &  9.59 $\pm$ 0.42 & 18.8 &  53699.7682      &           &2 / 3&\\
    &  &  & &  & & 27.366 $\pm$ 3.345e$^{-3}$ &  6.69 $\pm$  0.42 & 13.4    &  53699.7680         &  & &\\
        &  &  & &  & & 41.107 $\pm$ 6.306e$^{-3}$ &  3.19 $\pm$  0.32 & 6.2    &  53699.7687         &  & &\\

21 & 054.00014 & 07:56:31 & +21:52:29 &  9.50 &  9.74 & 20.310 $\pm$ 6.086e$^{-5}$ & 2.89 $\pm$ 0.14 & 8.5 &  53702.8142      &        A5 &9 / 19& NSV 3816\\
    &  &  & &  & & 45.725 $\pm$ 7.022e$^{-5}$ &  2.16 $\pm$  0.14 & 6.3    &  53702.8211         &  & &\\
        &  &  & &  & & 23.012 $\pm$ 7.570e$^{-5}$ &  2.32 $\pm$  0.14 & 6.8    &  53702.8130         &  & &\\

22 & 054.00075 & 07:58:27 & +21:36:24 & 10.80 & 11.11 & 22.591 $\pm$ 3.776e$^{-5}$ &  7.21 $\pm$ 0.026 & 12.9 & 53702.8148      &            &5 / 19&\\

23 & 064.00006 & 08:43:26 & +16:53:00 &  8.80 & 9.11  & 19.763 $\pm$ 3.791e$^{-1}$  & 6.84 $\pm$ 0.34 & 8.3 &  53812.5164      &           &1 / 2&\\

24 & 064.00050 & 08:43:45 & +17:25:00 & 10.60 & 10.94 & 20.148 $\pm$ 4.392e$^{-1}$ &  7.47 $\pm$ 0.44 & 8.6 & 53812.5238     &             &1 / 2&\\

25 & 066.00003 & 08:44:07 & +15:55:51 &  8.56 &  8.71 & 54.890 $\pm$ 9.400e$^{-1}$ &  6.37 $\pm$ 0.38 & 10.6 &  53774.6133       &       A0 &1 / 2&\\

26 & 062.00060 & 09:05:56 & +17:47:27 & 10.90 & 11.50 & 38.163 $\pm$ 8.580e$^{-4}$ &  11.24 $\pm$ 0.70 & 7.4 & 53753.6878      &         A5 &3 / 9&\\
&  &  & &  & & 46.333 $\pm$ 1.213e$^{-3}$ &  7.94 $\pm$  0.68 & 5.2    &  53753.7126         &  & &\\

27 & 062.00030 & 09:09:53 & +17:41:08 & 10.18 & 10.40 & 21.098 $\pm$ 3.614e$^{-5}$ &  7.64 $\pm$ 0.22 & 9.0 &  53753.6725     &            &4 / 9&\\

28 & 107.00023 & 13:10:39 & -06:25:01 & 10.25 & 10.51 & 20.214 $\pm$ 1.921e$^{-4}$   & 10.58 $\pm$ 0.30 & 10.4 &  53767.8800      &        A0 &4 / 10&\\

29 & 099.00087 & 15:58:40 & -19:27:14 & 11.62 & 12.46 & 19.673 $\pm$ 3.479e$^{-1}$ &  19.86 $\pm$ 0.92 & 12.4 & 53903.6396     &             &1 / 1&\\

30 & 148.00060 & 16:51:54 & +07:36:44 & 10.80 & 11.10 & 24.652 $\pm$ 4.632e$^{-4}$ & 27.71 $\pm$ 0.64 & 14.3 & 53812.7968   &            A5 &2 / 12&\\

31 & 012.00024 & 19:51:31 & -22:23:03 &  9.84 & 10.20 & 27.102 $\pm$ 7.279e$^{-1}$   & 8.22 $\pm$ 0.78 & 6.0 & 53959.6073    &        A6III &1 / 4&\\

32 & 153.00128 & 20:04:47 & -20:32:05 & 11.63 & 12.34  & 19.313 $\pm$ 5.077e$^{-1}$  & 43.72 $\pm$ 2.92 & 6.4 & 53919.7373    &              &1 / 1&\\

33 & 121.00214 & 21:01:43 & +16:39:37 & 11.20 & 11.57 & 23.471 $\pm$ 6.650e$^{-4}$ &  9.20 $\pm$ 0.48 & 6.2 & 53559.7694     &             &2 / 10&\\

34 & 121.00043 & 21:03:25 & +15:21:26 &  9.13 & 17.53 & 30.376 $\pm$ 1.866e$^{-3}$  & 3.61 $\pm$ 0.16 & 12.8 & 53750.7027    &              &3 / 10&\\

35 & 028.00439 & 21:54:03 & +25:11:07 & 12.85 & 13.54 & 19.445 $\pm$ 6.747$^{-5}$ &  32.02 $\pm$ 1.16 & 18.4 & 53575.7487       &           &5 / 9&\\
&  &  & &  & & 21.088 $\pm$ 1.405e$^{-4}$ &  15.39 $\pm$  1.02 & 8.8    &  53575.7599         &  & &\\
&  &  & &  & & 8.847 $\pm$ 1.238e$^{-3}$ &  9.79 $\pm$ 1.02 & 5.7    &  53575.7475         &  & &\\

36 & 028.01026 & 22:01:08 & +24:44:33 & 13.01 & 13.42  & 33.008 $\pm$ 5.843e$^{-5}$ &  36.45 $\pm$ 1.72 & 11.5 &  53575.7736     &            &5 / 9&\\

37 & 003.00147 & 22:01:53 & -12:28:52 & 12.24 & 12.23 & 17.731 $\pm$ 2.283e$^{-4}$ &  32.24 $\pm$ 1.14 & 18.5 & 53947.6845      &            &4 / 4&\\
&  &  & &  & & 41.446 $\pm$ 1.392e$^{-3}$ &  12.43 $\pm$ 1.28 & 7.4    &  53947.7181         &  & &\\
&  &  & &  & & 30.608 $\pm$ 8.642e$^{-4}$ &  9.26 $\pm$ 1.26 & 5.4    &  53947.7213         &  & &\\

38 & 138.00022 & 22:05:55 & +28:02:32 &  9.54 & 10.05 &17.767 $\pm$ 6.301e$^{-5}$ &  5.11 $\pm$ 0.20 & 12.0 &  53781.7689        &      A3 &6 / 13&\\
&  &  & &  & & 18.090 $\pm$ 9.286e$^{-5}$ &  3.50 $\pm$ 0.20 & 8.2    &  53781.7733         &  & &\\

39 & 138.00110 & 22:09:41 & +28:12:40 & 10.80 & 11.30 & 16.877 $\pm$ 1.751e$^{-4}$ &  7.87 $\pm$ 0.32 & 12.6 & 53584.7897    &              &4 / 13&\\

40 & 030.00019 & 22:59:23 & +37:14:33 &  9.26 &  9.65 & 20.199 $\pm$ 9.589e$^{-4}$ & 5.21 $\pm$ 0.20 & 13.5 & 53657.6049          &     F2 &4 / 8&\\
&  &  & &  & & 29.390 $\pm$ 1.564e$^{-3}$ &  2.60 $\pm$ 0.18 & 6.7    &  53657.6029         &  & &\\
&  &  & &  & & 19.757 $\pm$ 1.579e$^{-3}$ &  2.86 $\pm$ 0.20 & 7.4    &  53657.5702         &  & &\\

41 & 030.00195 & 23:02:06 & +36:30:28 & 11.51 & 11.91  & 22.278 $\pm$ 5.635e$^{-4}$   & 13.56 $\pm$ 0.42 & 21.4 & 53657.5695      &            &5 / 8&\\
&  &  & &  & & 18.802 $\pm$ 1.529e$^{-3}$ &  5.14 $\pm$ 0.40 & 8.1    &  53657.6120         &  & &\\

\end{longtable}
\end{center}

{\noindent} $^a$ Combination of TAOS field ID and TAOS star ID.\\
 $^b$ We doubled amplitudes derived by {\fontfamily{pcr}\selectfont PERIOD04}. \\
 $^c$ SNR of frequencies derived using {\fontfamily{pcr}\selectfont PERIOD04}. \\
 $^d$ The number of identifications / the number of zipper runs. Note that we did not count zipper runs observed by only one telescope.\\
 $^e$ The B8 star found to be an A5 star as explained in the text.\\
 $^f$ The G5 star found to be an F0 star as explained in the text.
 }}
\end{landscape}

\twocolumn

\label{lastpage}

\end{document}